\documentclass[prl,twocolumn,showpacs,floatfix,superscriptaddress]{revtex4}
\usepackage{graphicx,amsfonts,amssymb,amsmath,bm, hyperref}
\usepackage[applemac]{inputenc}

\newif\ifhyper
\hypertrue
\ifhyper
\hypersetup{
  citecolor = {green},
  colorlinks = {true}, % false
  urlcolor = {blue} % magenta
}
\fi

\newlength{\ldag}
\begin{document}

\title{Glassy  phase  in  quenched  disordered   crystalline membranes}

\author{O. Coquand} 
\email{coquand@lptmc.jussieu.fr}
\affiliation{Sorbonne Université, CNRS, Laboratoire de Physique Théorique de la Matière Condensée, LPTMC, F-75005 Paris, France}

\author{K. Essafi} 
\email{karim.essafi@oist.jp}
\affiliation{Okinawa Institute of Science and Technology Graduate University, Onna-son, Okinawa 904-0495, Japan}

\author{J.-P.  Kownacki}
\email{kownacki@u-cergy.fr}
\affiliation{LPTM, CNRS UMR 8089-Universit\'e de Cergy-Pontoise,   2 avenue Adolphe Chauvin, 95302 Cergy-Pontoise  Cedex, France}

\author{D. Mouhanna} 
\email{mouhanna@lptmc.jussieu.fr}
\affiliation{Sorbonne Université, CNRS, Laboratoire de Physique Théorique de la Matière Condensée, LPTMC, F-75005 Paris, France}

%------------------------------------------------------------------------------

\begin{abstract}

 We investigate  the flat phase  of $D$-dimensional crystalline  membranes embedded in a $d$-dimensional space and submitted to  both metric and curvature  quenched   disorders using  a nonperturbative renormalization group   approach. We  identify a second order phase transition controlled by  a  finite-temperature, finite-disorder fixed point  unreachable  within the  leading order of  $\epsilon=4-D$ and $1/d$ expansions. This critical  point  divides  the  flow diagram  into two  basins  of attraction:   that associated with  the  finite-temperature  fixed point controlling the long-distance behaviour of  disorder-free membranes and that associated with   the zero-temperature, finite-disorder  fixed point. Our work thus strongly suggests  the existence of a  whole  low-temperature glassy   phase  for  quenched disordered   crystalline membranes  and, possibly,  for graphene and  graphene-like compounds. 

\end{abstract}


\maketitle

{\sl Introduction.}  Graphene  \cite{novoselov04} is now well recognized as  a unique  material due to  its outstanding mechanical, optical, thermal, chemical  and electronic  properties \cite{castroneto09,katsnelson12,amorim16,akinwande17}:  high  mechanical   strength,   optical transmittance,  thermal conductivity and  carrier mobility.     These properties    make  it  one of the most  studied compounds both from   fundamental and  practical viewpoints.  At the fundamental level, graphene  has  launched   very challenging  problems, notably that  of unraveling the physics of Dirac massless fermions  propagating in a fluctuating curved space that mimics quantum gravity \cite{vozmediano10}.  At the  practical level,  graphene   has been regarded  as a very  promising candidate   for a broad   range of technological applications going from   energy storage,  ultrafiltration,   gas/electrochemical sensor  to  drug/gene delivery, bioimaging, chemo/bio sensing   and so on (see for instance \cite{sahu15}).     This situation  has   stimulated the design and  study   of other  two dimensional (2D)  materials such as  silicene, germanene, phosphorene, hexagonal boron  nitride or   transition metal dichalcogenides -- such as MoS$_2$ --   endowed   with  properties analogous  to those of graphene  except  for  the presence of a  nonvanishing or   tunable  band gap   making them, notably,    good candidates  for   the design of  semiconductor devices    \cite{akinwande17,roldan17}.

It is worth  recalling that the  outstanding    properties  of  graphene and, more generally,  graphene-like compounds   mainly  rely on the extreme purity and  regularity of   their  periodic  lattice structure. However,  as in all  materials,   lattice imperfections such as   defects (dislocations, grain boundaries, etc.),  impurities or vacancies  are either naturally present or  are  generated  during the manufacturing process.  These  defects or impurities     can dramatically  deteriorate    the  performances  of  pristine graphene, such as  its carrier mobility or thermal conductivity  (see, for instance, \cite{banhart11,Liu15}).  Conversely,    defective graphene has been  shown to display   enhanced sensing properties  and a  paradoxical increase  of its   elastic modulus  for  moderate  density of vacancies \cite{lopez15}.    Moreover, and of utmost importance,    the production  of vacancies,  introduction of   impurities, or  attachment of a chemical functional group have been  proposed as  one of the possible mechanisms to  open a tunable band gap. This {\sl  defect engineering}  of graphene  is, however, still at an early stage  as   even   the role   of  defects on its  mechanical properties   is still not fully understood. Also,  many properties   of  defective graphene seem to lack  universality as they  strongly rely on  the nature  and mobility of the  defects involved,  on the chemical functional group possibly attached to vacancy defects, etc.  \cite{banhart11,Liu15}. 

We  propose  here  a first step toward  the understanding   of  quenched disorder  in  graphene-like materials   by investigating    the long-distance effective action  of quenched disordered   crystalline membranes    (see \cite{proceedings89,bowick01} for reviews)   that   describes  the  elastic and curvature degrees of freedom  of  these systems.  The relevance   of our  predictions  to genuine graphene-like materials obviously relies on the robustness of our results with respect to the introduction of   electronic degrees of freedom, that are neglected here. We, however,  emphasize  the remarkable success of this kind  of  approach that  has already  successfully  explained   the  existence and stability   of  pristine  graphene  (see, for instance, \cite{katsnelson13}),  which is  {\it a priori}   forbidden     by  the Mermin-Wagner theorem,   by the emergence  of a  nontrivial  scaling of the physical  parameters in the deep infrared (IR), notably by   the infinitely growing  bending rigidity constant,   $\kappa(q)\sim q^{-\eta}$.     Accurate computations of the exponent  $\eta$   have  been realized  by various field-theoretical techniques --   perturbation theory \cite{nelson87,aronovitz88,guitter89}, self consistent screening approximation (SCSA)    \cite{,aronovitz89,ledoussal92,gazit09,zakharchenko10,roldan11,ledoussal17}, nonperturbative renormalization group (NPRG)  approaches \cite{kownacki09,braghin10,hasselmann11,essafi14} --   that compare very well to results obtained by  Monte-Carlo  and molecular dynamics simulations of graphene \cite{los09}.   Following up on this success we  consider  here  the   influence   of quenched disorder on the long-distance behaviour of  generic crystalline membranes and, hopefully, on graphene-like materials.  

 Largely motivated by the observation  of  a remarkable {\sl wrinkling  transition}   to a glassy  phase  upon cooling of  partially polymerized   phospholipid vesicles \cite{Mutz91,chaieb06,chaieb07},  quenched  disorder  has been  thoroughly  studied along these  lines  (for a review see  \cite{radzihovsky04}).   Nelson and Radzihovsky \cite{nelson91,radzihovsky91b}, using self-consistent techniques and  $\epsilon=4-D$ expansion,   have found that  purely  metric  disorder  was  irrelevant at any finite temperature $T$, the renormalization group (RG)  trajectories  being attracted toward   the   finite-$T$, vanishing-disorder  fixed point.   At vanishing  temperatures,   disorder would lead to a  destabilization of the flat phase through a disorder-induced softening  of the effective bending rigidity,  which has led  these authors  to   speculate about  the existence of   a spin-glass-like phase.   Soon thereafter Morse {\it et al.}  \cite{morse92a,morse92b},  extending   the work  done in  \cite{nelson91,radzihovsky91b}  by adding  curvature disorder,   have    demonstrated   the irrelevance of   this kind of  disorder  at any finite temperature. Moreover,   they have shown  that the  interplay between   metric  and curvature  quenched disorders   gives rise to a $T=0$ fixed point  {\it unstable} with respect to  temperature. Associated with  this fixed point, one  should  observe, at sufficiently low temperatures  and high values  of  disorder,   an anomalous disorder-induced   scaling regime \cite{morse92a,morse92b}.    These works  have been pursued by  an  explicit  search for  either flat-glassy  or crumpled-glassy phases  by means of mean-field approximations  in the case  of  short-range   \cite{radzihovsky92,bensimon92,bensimon93,attal93,park96} or    long-range  \cite{ledoussal93,mori94a,ledoussal17} disorders.

We have revisited  the model  considered in  \cite{morse92a,morse92b}  within a  NPRG  framework  in order  to go beyond the  early  $\epsilon$, {$1/d$}  expansions  and   SCSA approaches  and have obtained   a surprising  result:  in the presence of both  metric and curvature disorders, there exists  a finite-$T$, finite-disorder   fixed point   missed  within these  approaches.  This fixed point which is {\it unstable}  with respect to temperature  and,  thus, associated with  a second-order phase transition,  divides  the space of coupling constants into two basins of attraction:  one controlled by the finite-$T$, vanishing-disorder fixed point  associated with    disorder-free membranes and  another one   controlled by the $T=0$, finite-disorder  fixed point identified in  \cite{morse92a,morse92b} that  we now  find to be {\it stable}  within our approach. We thus predict   a whole ``glassy"  \footnote{We call  --  loosely  speaking -- this phase ``glassy"  in reference to  the fact that it  is  purely controlled  by disorder fluctuations.}     phase  controlled  by this  fixed point.

{\sl  Effective action.}   Let us consider a $D$-dimensional membrane embedded  in a $d$-dimensional space.  Each point of the membrane  is identified, within the membrane,  by   $D$ {\it internal} coordinates $\bm{ x}\equiv x_i$, $i=1\dots D$  and,  in the embedding space,  by  the $d$-dimensional vector  field  $\bm R(\bm x)$. The  long-distance, effective,  action which we consider is  given by  \footnote{We closely follow the notations of  Refs.\cite{morse92a,morse92b} in order to make comparisons easier. }: 
\begin{equation}
\begin{array}{ll}
S[\bm{ R}] =\displaystyle \int \text{d}^Dx \hspace{-0.3cm}  & \displaystyle \ \Big\{{\kappa \over 2}\big(\partial^2_{i}\bm{ R(x)}\big)^2 
+ {\lambda\over 2}\,  u_{ii}(\bm{ x})^2 +  {\mu}\,  u_{ij}(\bm{ x})^2 \\
& -{\bm c}(\bm{ x}). \partial_i^2\bm{ R(x)}   - \sigma_{ij}(\bm{ x}) \, u_{ij}(\bm{ x})\Big\}
\label{action}
\end{array}
\end{equation}
with summation over repeated indices.  In Eq.(\ref{action}) the first term represents   curvature  energy    with bending rigidity  $\kappa$,  and the second and third terms  the elastic energies    with Lamé coefficients $\lambda$ and $\mu$;  stability considerations require $\kappa$,  $\mu$,  and $\lambda+(2/D) \mu$ to be positive.    The  fourth and fifth terms represent couplings  of  disorder fields   ${\bm c}(\bm x)$ and  $\sigma_{ij}(\bm{ x})$  to the  curvature $\partial^2_{i}\bm{ R(x)}$ and  strain tensor $u_{ij}(\bm x)$, respectively. These  fields  are chosen  to be short-ranged quenched Gaussian ones  with  zero-mean value and variances given by  \cite{morse92a,morse92b}: 
\begin{equation}
\begin{array}{ll}
[c_{i}(\bm{ x})\  c_{j}(\bm{ x}')]= \Delta_{\kappa}\,  \delta _{ij} \ \delta^{(D)}(\bm{ x-x'})
\\
\\
\big[\sigma_{ij}(\bm{ x})\   \sigma_{kl}(\bm{ x'})\big]= (\Delta_{\lambda} \delta _{ij} \delta_{kl}+2 \Delta_{\mu} I_{ijkl})\ \delta^{(D)}(\bm{ x-x'})
\label{variance}
\end{array}
\end{equation}
where $I_{ijkl}={1\over 2}(\delta_{ik}\delta_{jl}+\delta_{il}\delta_{jk})$, with $i,j,k,l=1\dots D$,  $[\dots]$ denotes an average over Gaussian disorder and  where $\Delta_{\kappa}$,  $\Delta_{\mu}$ and $\Delta_{\lambda}+(2/D) \Delta_{\mu}$ are positive.   For disorder-free  membranes,   the strain tensor $u_{ij}$  in Eq.(\ref{action})   is given by: $u_{ij}={1\over 2}(g_{ij}-g_{ij}^0)={1\over 2}(\partial_i \bm{ R}.\partial_j\bm{ R}- \partial_i \bm{ R}^0.\partial_j\bm{ R}^0)$ where $g_{ij}$  represents a metric on the membrane and 
\begin{equation}
\bm{ R}^0(\bm{ x)}=[\langle \bm{ R(\bm{ x})}\rangle]=\zeta  x_i \bm{ e}_i 
\label{flat}
\end{equation}
in which     $\langle\dots\rangle$  denotes   a thermal average.  In Eq.(\ref{flat}) the $\bm{ e}_i$ form  an orthonormal set of $D$ vectors  so that   $\bm{ R}^0(\bm{ x)}$   represents  a flat configuration of reference.     In this  configuration   the tangent vectors $\partial_i \bm{ R}^0$  take   nonvanishing values,   $\partial_i \bm{ R}^0=\zeta \bm{ e}_i$,  where $\zeta$  is  the  stretching factor, so that  $g_{ij}^0$ is just the flat metric:  $g_{ij}^0=\zeta^2 \delta_{ij}$.   In the presence of disorder, the reference  configuration is no longer the flat one. Quenched random fields  induce \cite{morse92a,morse92b} {\it  i)} a local deformation of  $g_{ij}^0$, noted  $\delta g_{ij}(\bm{ x})$,   given by : $\sigma_{ij}(\bm{ x})={1/2}(\lambda\delta_{ij}\delta_{kl}+2 \mu I_{ijkl})  \delta g_{kl}(\bm{ x})$  and  {\it  ii)}  a  random spontaneous curvature: $\partial^2_{i}{\bm R}(x)={\bm c}(\bm{ x})/ \kappa$.

 The  action Eq.(\ref{action})  is very close to that considered   in  \cite{morse92a,morse92b}. However ours  differs  from the latter one  by the fact  that  it is entirely written in terms of the  field $\bm{R}$  instead of    the fields $\bm{u}$ and $\bm{h}$ that generally parametrize fluctuations around the flat phase.  This  allows one to keep a fully rotationally invariant formalism -- where,  notably,  the full strain tensor $u_{ij}$  is considered without any approximation -- in which    the crumpling-to-flat transition can be investigated as well \cite{kownacki09,braghin10,essafi11,hasselmann11,essafi14}.   The flat phase is obtained   through  the   limit  where the  running stretching factor  $\zeta_k$ -- see below --  that signals the appearance of a spontaneous symmetry breaking,  takes a finite value or, equivalently,  where  the dimensionless  stretching  factor   $\overline\zeta_k$ goes to infinity (see for instance  \cite{coquand16a}).

{\sl  NPRG approach.}   To  derive the RG equations  for the coupling constants entering in action Eq.(\ref{action})  we first  average over quenched disorder by means of  replica formalism.    Then we employ  a method based on the concept of effective average action  $\Gamma_k$ -- where $k$ is a running scale  --  analog to a Gibbs free energy where only fluctuations of momenta $q\ge k$ have  been  integrated out  (see  \cite{wetterich93} and \cite{berges02,delamotte03,pawlowski07,kopietz10,gies12,delamotte12,rosten12,nagy14} for reviews and, for recent applications to frustrated magnets \cite{delamotte16,dupuis16a,dupuis16b}, turbulence \cite{canet16}, Kardar–Parisi–Zhang equation \cite{canet11a,canet11b} or random fields models \cite{tarjus04,tarjus06,tarjus08a,tarjus08b,mouhanna10,mouhanna16}).  At the microscopic lattice scale $\Lambda\propto a^{-1}$ where $a$ is the typical length of a chemical bound, $\Gamma_{k=\Lambda}$ coincides with the microscopic action $S[\bm{ R}]$ while, at long distance, $k=0$,   it coincides with the usual Gibbs free energy $\Gamma[\bm{ R}]$. The  effective average action $\Gamma_k$ follows an   exact equation \cite{wetterich93c}: 
\begin{equation}
\partial_t \Gamma_k[\bm{ R}]={1\over 2} \hbox{Tr} \Big\{{\partial_t  R_k}\,  (\Gamma_k^{(2)}[\bm{ R}]+R_k)^{-1}\Big\}
 \label{renorm}
\end{equation}
where   $t=\ln \displaystyle {k / \Lambda}$ and where  the trace must  be understood as   a  space (or momentum)  $D$-dimensional integral   as well as a summation over  implicit   vectorial and replica indices. In Eq.(\ref{renorm}),    $\Gamma_{k,  ij}^{\alpha \beta\, (2)}[\bm{ R}]$  is the  inverse propagator, the second derivative of $\Gamma_k$ with respect to the field $\bm{R}$. Finally,   $R_k(\bm{q})$  is  a  cut-off function that suppresses  the propagation of modes with momenta $q<k$ and makes   that $\Gamma_k$ encodes only modes with momenta $q> k$ \cite{berges02}.  We  consider a cut-off  function  diagonal both in vector and replica space -- $R_{k,{ij}}^{\alpha\beta}(q)=R_k(q)\delta_{ij}\delta_{\alpha\beta}$ -- and we  present  our   analytical  results  for the "$\Theta$" cut-off \cite{litim00},   $R_{k}(q)\propto (k^4-q^4)\Theta(k^2-q^2)$.  Equation (\ref{renorm}) is  a functional  partial differential equation  that cannot be solved  exactly  so that   approximations are required. We use here  the field, field-derivative expansion  \cite{berges02} where  $ \Gamma_k[\bm{ R}]$   is expanded in powers of  the order parameter $\partial_i \bm{ R}$ and its derivatives around the nontrivial minimum, Eq.(\ref{flat}),  {\it while}  preserving  the nonperturbative content of Eq.(\ref{renorm}). We are thus led  to the following effective action:
\begin{equation}
\begin{array}{ll}
 \hspace{-0.2cm}\Gamma_k[\bm{ R}] =\displaystyle  \int & \text{d}^Dx \  \bigg\{ \displaystyle {Z_k \over 2}\big(\partial_{i}^2\bm{ R}^{\alpha} \big)^2 - {\widetilde\Delta_{\kappa k}\over 2}  \partial_{i}^2\bm{ R}^{\alpha}. \partial_{j}^2\bm{ R}^{\beta} J^{\alpha\beta}
\\
& \hspace{-2cm} \displaystyle +    {\widetilde\lambda_k\over 8} \big(\partial_{i}\bm{ R}^{\alpha}.\partial_{i}\bm{ R}^{\alpha}-D {\tilde\zeta_k}^2\big)^2+ \displaystyle   {\widetilde\mu_k\over 4} \big(\partial_{i}\bm{ R}^{\alpha}.\partial_{j} \bm{ R}^{\alpha}-\delta_{ij}  \tilde\zeta_k^2\big)^2  
\\
&  \hspace{-2cm}   \displaystyle  -    {\widetilde\Delta_{\lambda k}\over 8} \big(\partial_{i}\bm{ R}^{\alpha}. \partial_{i}\bm{ R}^{\alpha}- D \tilde\zeta_k^2 \big) \big( \partial_{j}\bm{ R}^{\beta}. \partial_{j}\bm{ R}^{\beta}-D \tilde\zeta_k^2 \big)
\\
&\hspace{-2cm}  \displaystyle   -  {\widetilde\Delta_{\mu k}\over 4}  \big( \partial_{i}\bm{ R}^{\alpha}. \partial_{j}\bm{ R}^{\alpha}-  \delta_{ij}\tilde \zeta_k^2\big) \big( \partial_{i}\bm{ R}^{\beta}. \partial_{j}\bm{ R}^{\beta}- \delta_{ij}\tilde \zeta_k^2 \big)\bigg\}\ . 
\label{effectiveaction2}
\end{array}
\end{equation}
where $J^{\alpha\beta}$  is the  unit matrix  in replica space.  Note that although limited to  fourth order    in powers of the order parameter $\partial_{i}\bm{ R}$  and to first order   in powers of   its derivatives, this  truncation  has  led to very  well converged  results for pure membranes, (see \cite{kownacki09,braghin10,hasselmann11,essafi14}), leading in particular to the value of $\eta\simeq 0.85$  perfectly confirmed numerically  \cite{los09}.         In Eq.(\ref{effectiveaction2}) we have done the rescaling  $\bm{ R}\mapsto  T^{1/2} Z_k^{1/2} \kappa^{-1/2} \bm{ R}$, where $Z_k$ is a  running  field renormalization   and  introduced  the running coupling constants:    $\widetilde \lambda_k=\lambda TZ_k^2 \kappa^{-2}$,  $\widetilde \mu_k=\mu TZ_k^2 \kappa^{-2}$,   $\widetilde\Delta_{\lambda k}=\Delta_\lambda Z_k^2\kappa^{-2}$, $\widetilde\Delta_{\mu k}=\Delta_\mu Z_k^2\kappa^{-2}$,  $\widetilde\Delta_{\kappa k}=\Delta_\kappa T^{-1}Z_k\kappa^{-1} $  and $\tilde \zeta_k=\zeta_k T^{-1/2}  Z_k^{-1/2}  \kappa^{1/2}$.  Note that  $\widetilde \mu_k$  and  $\widetilde \lambda_k$   can be used as  a  measure  of the temperature $T$   while $\widetilde\Delta_{\kappa k}$ diverges at vanishing temperatures. In order to study this latter  regime we introduce, as in \cite{morse92a,morse92b},   $\widetilde g_{\mu k}=\widetilde \mu_k  \widetilde\Delta_{\kappa k}$ and $\widetilde g_{\lambda k}=\widetilde \lambda_k  \widetilde\Delta_{\kappa k}$  that stay  finite at any  temperature.  The field  renormalization $Z_k$  and  rescaled curvature disorder  variance $\widetilde\Delta_{\kappa k}$ allow  to define the   running anomalous dimensions    $\eta_k=-\partial_t \ln Z_k$  and  $\eta_k'=  \eta_k+\partial_t \ln \widetilde\Delta_{\kappa k}$  that, at a fixed point,  characterize  the scaling behaviour of the thermal  -- $\chi({\bm q})$ -- and disorder --  $C({\bm q})$ -- correlation functions, respectively,  that are defined  from the two-point correlation function $G_{{\bm R}{\bm R}}({\bm q})= \big[\langle {\bm R}({\bm q})\bm{R}(-{\bm q})\rangle \big]$ by:
\begin{equation}
\begin{array}{ll}
G_{{\bm R}{\bm R}}({\bm q})&= \big[\langle \delta {\bm R}({\bm q})  \delta \bm{R}(-{\bm q})\rangle \big]+\big[\langle {\bm R}({\bm q})\rangle \langle \bm{R}(-{\bm q})\rangle \big]\\
\\
&= T \chi({\bm q})+ C({\bm q})
\label{correlation}
\end{array}
\end{equation}
with $\delta {\bm R}({\bm q})={\bm R}({\bm q})-\langle {\bm R}({\bm q}) \rangle$,  and  behave, at low momenta,  as:  
\begin{equation}
\chi({\bm q})\sim q^{-(4-\eta)}, \hspace{0.5cm}  C({\bm q})\sim  q^{-(4-\eta')}\ .
\end{equation}
 Finally,  we define from  $\eta$ and $\eta'$ the   exponent  $\phi$ by \cite{morse92a,morse92b}: $\phi=\eta'-\eta$ that describes, in particular,  the flow of the  temperature near  $T=0$. 
 
{\sl  RG equations.}  We consider the RG equations obtained  from Eq.(\ref{effectiveaction2}) in the flat phase.    We  restrict our study to the attractive hypersurface   defined by  $\widetilde{\Delta}_{\lambda k}=-2/D_2\, \widetilde{\Delta}_{\mu k}$ and $\widetilde{g}_{\lambda k}=-2/D_2\, \widetilde{g}_{\mu k}$  -- where we have  introduced  the notation $D_n=D+n$ for $n>0$ -- that  generalizes to any dimension $D$  the one  considered near  $D=4$ in \cite{nelson91,radzihovsky91b,morse92a,morse92b} (see also  \footnote{We indicate that only in the case  $D<4$ we find it  attractive in the IR.}).   The  RG equations, in terms of dimensionless coupling  constants,   $\overline\mu_k=Z_k^{-2}\, k^{D-4}\widetilde\mu_k$,    $\overline\Delta_{\mu k}=Z_k^{-2}\, k^{D-4}\widetilde\Delta_{\mu k}$, $\overline\Delta_{\kappa k}=Z_k^{-1}\, \widetilde\Delta_{\kappa k}$    and    $\overline g_{\mu k}=Z_k^{-3}\, \, k^{D-4}  \widetilde g_{\mu k}$ read: 
\begin{equation}
\begin{array}{ll}
 \partial_t\overline{\mu}_k=(D-4+2 \eta_k)\overline{\mu}_k+4 d_c\, \overline{\mu}_k  ( \overline{\mu}_k\, \tilde A_D+3\, \overline g_{\mu k}\,   \breve{A}_D )
\\
\\
  \partial_t\overline{\Delta}_{\mu k}=(D-4+2 \eta_k)\overline{\Delta}_{\mu k}  \\
  \\
 \hspace{1.1cm} + 8 d_c  \big( \overline{\Delta}_{\mu k}\,  \overline\mu_k\, \tilde A_D+\overline g_{\mu k}(3\,  \overline{\Delta}_{\mu k}\,  \breve{A}_D- \overline g_{\mu k}  \Acute{A}_D)\big)
 \\
\\
\partial_t\overline{\Delta}_{\kappa k}=\overline{\Delta}_{\kappa k}\big(\eta_k-4  D(D-1)  \overline{\Delta}_{\mu k} \tilde A_D\big)
\\
\\
\partial_t\overline g_{\mu k}= (D-4+3 \eta_k)\overline g_{\mu k}  \\
\\
 \hspace{0.5cm} - 4\overline g_{\mu k}\big(D(D-1)  \overline{\Delta}_{\mu k} \tilde A_D -d_c(\overline{\mu}_k \tilde A_D+3 \overline g_{\mu  k}\breve{A}_D)\big)
\label{eqrg}
\end{array}
\end{equation}
where  $\eta_k$  is given by: 
\begin{equation}
\eta_k= {32(D-1)D_8\big(2 D \overline g_{\mu k}+D_4(\overline\mu_k -\overline\Delta_{\mu k})\big) A_D\over  D D_2 D_4 D_8+32(D-1)\big(2 D \overline g_{\mu k}+D_8(\overline\mu_k-\overline\Delta_{\mu k})\big)A_D}
\label{eta}
\end{equation}
and  
\begin{equation}
\eta'_k= 2 \eta_k -4  D(D-1)  \overline{\Delta}_{\mu k} \tilde A_D
\label{etaprime}
\end{equation}
where   $\tilde A_D=16 A_D(D_8-\eta_k)/(D D_2 D_4 D_8)$,    $\breve{A}_D=16 A_D(D_{12}-\eta_k)/(D D_2 D_8 D_{12})$,  $\Acute{A}_D=16 A_D(D_{16}-\eta_k)/(D D_2 D_{12} D_{16})$, $A_D^{-1}=2^{D+1}\pi^{D/2}\,\Gamma(D/2)$,   $\Gamma(\dots)$ being   Euler's gamma function and  $d_c=d-D$.
The RG Eqs.(\ref{eqrg})-(\ref{etaprime}) generalize those derived perturbatively  in \cite{nelson91,radzihovsky91b,morse92a,morse92b} with the major difference that our expressions of    $\eta_k$ and $\eta_k'$  involve   nonpolynomial, thus nonperturbative,  contributions  of the disorder   that  play  a crucial role, see below.  

{\sl $\epsilon$-expansion.} Close to  $D=4$, expanding  our equations in powers of the coupling  constants  and in $\epsilon=4-D$   we  recover  \footnote{Up to  a redefinition of the coupling constants by a factor $1/8\pi^2$.}  those  obtained in \cite{morse92a,morse92b}.
%begin{equation}
%\begin{array}{ll}
%\partial_t{\overline\mu_k}=\displaystyle -\epsilon{\overline\mu_k} +{\overline\mu_k\over 96 \pi^2} \Big(24 \overline{\Delta}_{\mu k}+2(12+d_c)\overline g_{\mu k} \\
%\\ \hspace{4cm}+  (24+d_c)\overline \mu_k\Big)
%\\
%\\
% \partial_t\overline g_{\mu k}=\displaystyle -\epsilon\overline g_{\mu k}+  {\overline g_{\mu k}  \over 96 \pi^2}\Big(-48\overline{\Delta}_{\mu k}+2(18+d_c)\overline g_{\mu k} \\
% \\
% \hspace{4cm} + (36+d_c)\overline \mu_k\Big)
% \\
% \\
% \partial_t\overline{\Delta}_{\mu k}=\displaystyle  -\epsilon \overline{\Delta}_{\mu k} +{ \overline{\Delta}_{\mu k} \over 48 \pi^2}\Big(-12\overline{\Delta}_{\mu k}+2(6+d_c)\overline g_{\mu k}\\
%\hspace{4cm}  +\displaystyle (12+d_c)\overline \mu_k\Big)-  d_c {\overline g_{\mu k}^2\over 96\pi^2}
%\label{eqrg4D}
%\end{array}
%\end{equation}
%and $\eta_k={1/ 8\pi^2}(\overline\mu_k + \overline g_{\mu k}-\overline\Delta_{\mu k})$. 
Let us recall their  content  in  $D<4$.   There  exists   one {\it  fully stable}  fixed point, called $P_4$,   lying at  finite temperature -- $\overline\mu_4= 96 \pi^2  \epsilon/(24 +d_c)$  --  characterized by  vanishing disorder coupling constants $\overline\Delta_{\mu 4}$, $\overline g_{\mu 4}$ and by   $\eta_4=\eta_4'/2=12 \epsilon/(24 +d_c)$; it is  thus  associated with  disorder-free membranes.  There  exists another  fixed point, named $P_5$, lying at  vanishing temperature     -- $\overline\mu_5=0$ --   characterized by  nonvanishing disorder  coupling constants  $\overline{\Delta}_{\mu 5}=24 \pi^2  \epsilon/(6+d_c)$,  $\overline g_{\mu 5}=48 \pi^2  \epsilon/(6+d_c)$ and by $\eta_5=\eta'_5=\displaystyle{3\epsilon/(6+d_c)}$, {\it i.e.} $\phi_5=0$ at order $\epsilon$.    A further  stability analysis   \cite{morse92a,morse92b} shows  that $P_5$  is  marginally  {\it unstable}  with respect to  temperature so that   $ P_5$   should   be relevant to the physics of  membranes only  up to a typical lengthscale $L_c\sim e^{1/\alpha T}$ where  $\alpha$  depends on the  membrane parameters  \cite{morse92a,morse92b}.

{\sl Nonperturbative analysis.}  Studying now directly  the nonperturbative equations (\ref{eqrg})-(\ref{etaprime})  we are   led to a substantively  different conclusion.  First,   we   well identify, as in \cite{morse92a,morse92b},  the  {\it fully stable}  finite-$T$, vanishing-disorder fixed point $P_4$  and  the  $T=0$, finite-disorder   fixed point  $P_5$.  {\it However},  we find that  $P_5$   is, contrary to what  has been found in  \cite{morse92a,morse92b}, {\it stable} with respect to temperature,  see Fig.({\ref{Figflow}).  This situation  relies on the fact that  there exists   a supplementary  -- critical --  fixed point,  that we call $P_c$,   lying between $P_4$ and $P_5$, {\it unstable} with respect to temperature,   and which governs a second order phase transition  between the disorder-free and  the ``glassy" phases  associated with $P_4$ and $P_5$, respectively, see Fig.({\ref{Figflow})}. It is  very  instructive to derive from Eqs.(\ref{eqrg})-(\ref{etaprime})  the coordinates of $P_c$,  the  critical  exponent  $\eta_c$ and  the eigenvalue $y_c$  associated with the relevant direction near   $D=4$.  At  {\it leading nontrivial order in $\epsilon$}   one has  \footnote{All the quantities of order $\epsilon^2$ given here  are  approximative since  finite-order derivative approximations   of the NPRG approach are  not exact at order $\epsilon^2$ \cite{papenbrock95}.}  : $\overline\mu_c=4 \pi^2 \epsilon^2(27+5d_c)/15(6+d_c)^2$, $\overline{\Delta}_{\mu c}=24 \pi^2  \epsilon/(6+d_c)$,  $\overline g_{\mu c}=48 \pi^2  \epsilon/(6+d_c)$,  $\eta_c=\displaystyle{3\epsilon/(6+d_c)}$ which  equals  $\eta'_c$ since  $\partial_t \ln \widetilde\Delta_{\kappa k}$ vanishes at  $P_c$ and, finally,  $y_c=\displaystyle{(27+5d_c)d_c\epsilon^2/ 20(6+d_c)^3}$.   These quantities  are,  {\it at  first  order  in}  $\epsilon$,  strictly  equal  to  those associated with   $P_5$.   Therefore  $P_c$   cannot  be distinguished, at this order,  from $P_5$; this is the reason why it has been missed in \cite{morse92a,morse92b} while it is clearly identified within our approach, already  at order $\epsilon^2$  -- via  $\overline\mu_c$ and $y_c$  -- and, even more  clearly,   in the physical -- $D=2$ and $d=3$   -- case, see Fig.({\ref{Figflow})}.  To complete our results we provide in Table \ref{table}  the critical exponents $\eta$, $\eta'$  and the lower critical dimensions $D_{lc}$ associated with   $P_5$ and $P_c$  in the physical case both for the NPRG and SCSA  \footnote{For numerical results  we have   employed  two   families of  cut-off functions  $\widetilde R_k(q)=C\,  Z_k q^4 /  (\hbox{exp}(q^4/k^4)-1)$  and  $\widetilde R_k(q)=C\,  Z_k k^4  \hbox{exp}(-q^4/k^4)$ where  $C$  is a free parameter used   to investigate the cut-off dependence of physical quantities. Varying $C$ allows one to optimize each cut-off function inside its family, {\it  i.e.}  to (try to) find stationary values of these quantities, see for instance \cite{canet03a}.}. While the   quantitative agreement   at $P_5$   is remarkable as for $\eta_5$ and $D_{lc5}$,  what   firmly validates our approach,   a decisive discrepancy  between the NPRG and SCSA   is  that  the  fixed point $P_c$ is present within  the former  and  absent within the latter.  This also  results  in a disagreement  concerning  the stability of $P_5$, already observed at order $\epsilon^2$ where we find  $\phi_5= -y_c$ while the SCSA leads to  $\phi_5=0$. The origin of this discrepancy     is less trivial to identify than that occurring between the perturbative  (including leading order   of $1/d$ expansion)  and nonperturbative approaches because of the complexity of the set of approximations -- notably partial resummation  of infinite amount  of diagrams --  performed within the SCSA \cite{gazit09}.   This is under investigations \cite{coquand17}. 

\begin{table}[htbp]
\begin{tabular}{|c|c|c|c|c|c|c|c|c|c|}
\hline
 \hspace{0.5cm}  &  $\eta_5$ & $\eta'_5$ &   $\phi_5$   &  $D_{lc5}$  & $\eta_c=\eta'_c$  &  $y_c$ &  $D_{lcc}$  \\
\hline
\hline
SCSA \cite{ledoussal17} &  0.449  &0.449 & 0   &  1.70  &  $\times$   &  $\times$  & $\times$ \\
\hline 
  NPRG   &    0.449  & 0.277 &  $-0.172$ &  1.71  &   0.492    &  0.131  &  1.67  \\
\hline
\end{tabular}
\caption{The critical exponent  $\eta$, $\eta'$  and $D_{lc}$ at the fixed points $P_5$ and $P_c$.} 
\label{table}
\end{table}

%*********************************************************************************************************************************************
\begin{figure}[h]
\includegraphics[scale=0.32]{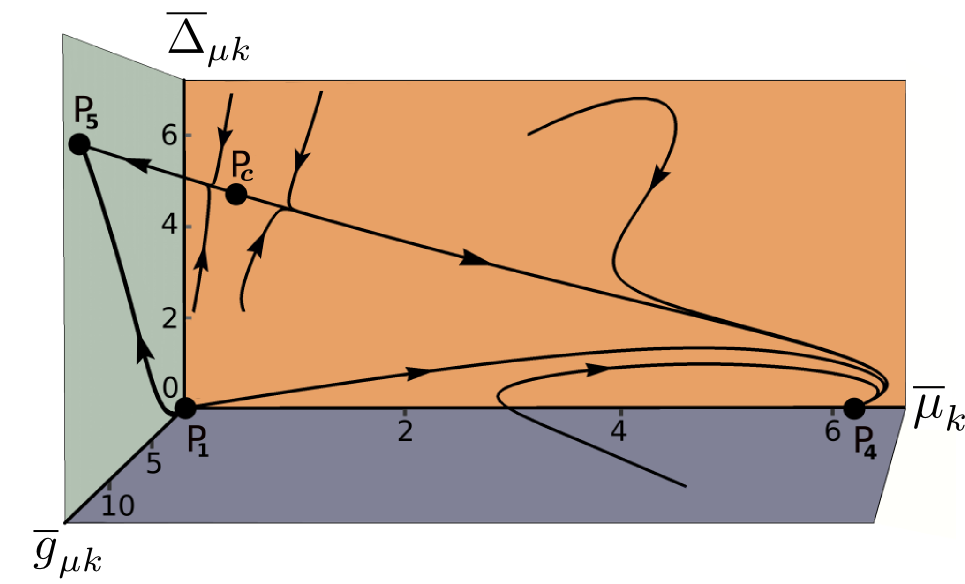}
\caption{The RG flow in $D=2$ and $d=3$ within the space $\{\overline\mu_k,\overline g_{\mu k},\overline{\Delta}_{\mu k},\overline{\Delta}_{\lambda k}=-1/2\, \overline{\Delta}_{\mu k}, \overline{g}_{\lambda k}=-1/2\, \overline{g}_{\mu k}\}$.  $P_1$ is the Gaussian, unstable, fixed point. $P_4$  is fully attractive and associated  with disorder-free membranes.  $P_5$   is also fully attractive and controls  the low-temperature phase of disordered membranes. Finally,   $P_c$, unstable with respect to temperature governs the phase transition between the two phases.}
\label{Figflow}
\end{figure}

%*********************************************************************************************************************************************

{\sl  Conclusion.}  Investigations  of  the nonperturbative regime of  quenched disordered  crystalline membranes  have revealed the existence of a second order phase transition between  a  disorder-free phase and a  glassy  phase  controlled by the  $T=0$  fixed point.  Several issues should be settled in the near  future.  First,  the question  of the   numerical  and experimental  probe  for  this phase  should  be  addressed.  Thanks to its potentially wide range of applicability, our prediction  could {\it a priori}  be tested on a large variety of systems.  In the context of     biological physics, cell membranes  with    imperfectly   polymerized  cytoskeleton -- inducing metric  disorder -- and with inclusion of asymmetrical proteins  -- generating  curvature disorder -- are  possible candidates. In the context of $2D$ electronic crystalline membranes,  an obvious candidate, among others,  is  graphene  in which   inclusion of  lattice defects   can  induce, in addition to metric alterations,  a  rearrangement of  sp$^2$-hybridized carbon atoms into nonhexagonal and, thus, nonvanishing curvature   structures. Second, at the formal level,   the very nature  of the  putative glassy  phase     should be  clarified \cite{coquand17}.   Also, more specifically in the context of graphene-like membranes,   the remarkable coupling  between electronic and elastic degrees of freedom   raises   both theoretically and experimentally  very challenging  issues  of  understanding  how  electronic, transport,  thermal, and  optical  properties   are altered  within  the  expected glassy  phase.

\section{Acknowledgements}

We thank  B. Delamotte, J.-N. Fuchs, P. Le Doussal, L. Radzihovsky,  G. Tarjus and M. Tissier  for helpful discussions. 

%\bibliography{bibliotheque1.bib}

\end{document}